\pdfoutput=1
\documentclass[letterpaper,10pt]{article}
\usepackage{osameet2}

\usepackage{amsmath,amssymb}

\usepackage[pdftex,colorlinks=true,bookmarks=false,citecolor=blue,urlcolor=blue]{hyperref} 

\begin{document}

\title{Machine learning-based Raman amplifier design}
\author{D. Zibar$^{1}$, A. Ferrari$^2$, V. Curri$^2$ and A. Carena$^2$}
\address{
1.~DTU Fotonik, Technical University of Denmark, DK-2800, Kgs.~Lyngby, Denmark\\
2.~DET, Politecnico di Torino, Corso Duca degli Abruzzi, 24 - 10129, Torino, Italy}
\email{dazi@fotonik.dtu.dk}

\begin{abstract}
A multi-layer neural network is employed to learn the mapping between Raman gain profile and pump powers and wavelengths. The learned model predicts with high-accuracy, low-latency and low-complexity the pumping setup for any gain profile.
\end{abstract}
\vspace{-0.5cm}

\section{Introduction}
Optical amplification schemes exploiting the stimulated Raman scattering (SRS) are an attractive solution for future multi-band optical communication systems, as SRS is a physical effect of silica, available independently of the deployed bandwidth \cite{Napoli18}.
Compared to the well-established Erbium doped fibre amplifiers (EDFAs), Raman amplifiers (RAs) offer low noise properties, because of distributed amplification, and gain availability across a broad range of wavelengths, when operated in multi-pump configurations. Moreover, RAs allow a flexible gain profile design.\vspace{0.1cm}     

The challenge with Raman amplifier design is on the selection of pump powers and wavelengths that would result in a targeted gain profile. In general, this is a complex optimization problem as it requires the solution 
of a two-point boundary condition problem.
Several approaches to pump powers and wavelengths selection have been reported in the literature \cite{OE_Neto_2007,Ferreira11} and references therein. The presented approaches in \cite{OE_Neto_2007,Ferreira11} involve integration of propagation equations describing multi-pump and optical Wavelength Division Multiplexed (WDM) channels interaction which is complex, highly time consuming and in some cases the solution may not converge \cite{OE_Neto_2007,Ferreira11}.\vspace{0.1cm}   

In this paper, we present a novel approach, based on machine learning, for solving the optimization problem of determining the pump powers and wavelengths resulting in a targeted gain profile. A multi-layer neural network is employed to learn the mapping between the gain profile and pump powers and wavelengths. Once the model has been learned, given a set of arbitrary gain profiles, pumps powers and wavelengths can be predicted by simple forward propagation through the multi-layer neural network. This solution offers a high-degree of flexibility compared to the state-of-the-art methods where for each gain profile complex optimization problem needs to be solved from scratch. Finally, the predicted pump powers and wavelengths result in gain profiles that have maximum error of up to 0.6 dB compared to the targeted gain profiles.
\vspace{0.1cm}   

The proposed approach has thus high-accuracy, low complexity and it is ultra-fast as the integration of propagation equation is avoided. This makes it highly attractive for application in network control-plane where almost real-time adjustments can be required and a full featured optimization process can not take place.  

\section{Machine learning framework}
\label{sec:theory}
\begin{figure*}[t]
{\includegraphics[width=\textwidth]{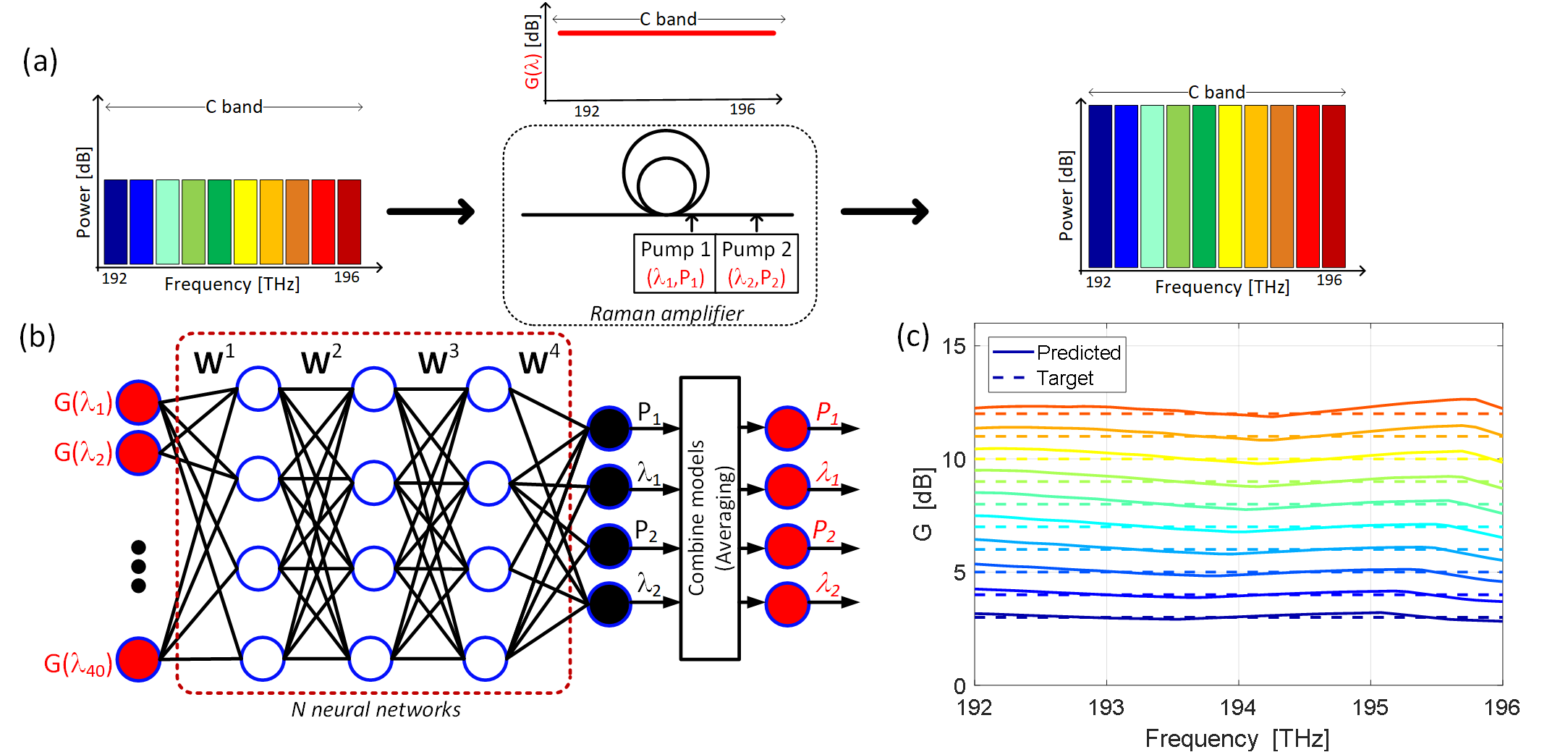}}
\caption{\small{(a) A single span Raman amplifier employing backward pumping with two pumps. A target gain profile is given and the objective is to determine the corresponding pump powers and wavelengths. (b) Multi-layer neural network employing model averaging for learning the mapping between Raman gain profile and pump powers and wavelength. The learned mapping is then used to predict pump powers and wavelengths settings for a targeted Raman gain profile. (c) \textit{Evaluation:} For the predicted pump powers and wavelengths, Raman solver is run to obtain the predicted gain profiles which are then compared to target gain profiles.}}
\label{fig:system_set_up}
\end{figure*}
In Fig.~\ref{fig:system_set_up}(a), a single span Raman amplification scheme is shown. The considered single span can be a part of a multi-span system and also Raman amplification can take place in conjunction with EDFA. The gain $G(\lambda)$ over the entire C-band (4 THz band between 192 and 196 THz) and backward Raman pumping scheme with two pumps are considered. Only two pumps are considered to keep the system simple, and also to demonstrate the full potential of the proposed method in delivering the required gain profile using a limited number of pumps over a large bandwidth. \vspace{0.1cm}

The objective is to determine pump powers and wavelengths settings that would result in specific gain profiles $G(\lambda)$ such as: tilted gain profile, flat gain profile or an arbitrary gain profile. For example, the targeted gain profile could be the one needed to compensate for EDFA gain ripples or tilt when RA is used a stage in hybrid amplifiers.

To determine the allocation of pump powers and wavelengths for a targeted gain profile, we first use a multi-layer neural network to learn the relationship between gain profile $\mathbf{G}=[G(\lambda_1),...,G(\lambda_{40})]$ at specified wavelengths and the pumps powers and wavelengths $[P_1,\lambda_1,P_2,\lambda_2]$, as illustrated in Fig.~\ref{fig:system_set_up}(b).  The second order optimization algorithm (Levenberg-Marquardt back propagation) is used to learn the weight matrices $[\mathbf{W}^1,\mathbf{W}^2,\mathbf{W}^3,\mathbf{W}^4]$ that connect input layer to hidden layers, and the hidden layers to the output. Once the learning has converged, the multi-layer neural network operates in a prediction mode where the desired pump powers and wavelengths are obtained by simple forward propagation of the gain profile $\mathbf{G}=[G(\lambda_1),...,G(\lambda_{40})]$ through the network with learned matrices $[\mathbf{W}^1,\mathbf{W}^2,\mathbf{W}^3,\mathbf{W}^4]$ and specified activation function ($\tanh$ in the considered case). To improve the learning in terms of the Mean Squared Error (MSE), we employ model combination. In simple word this means that we train $N=10$ independent neural networks in parallel and combine their outputs. For each run, the data set is randomly shuffled to emulate a new data set. \vspace{0.1cm} 

The data-set.~i.e.~$\mathcal{D}=\{G^i(\lambda_1),...,G^i(\lambda_{40}),P^i_1,P^i_2,\lambda^i_1,\lambda^i_2)|i=1,...,M\}$ used for the training is of relatively small size $M=2000$. The data set is obtained by running a full Raman solver for the Raman amplifier shown in Fig.~\ref{fig:system_set_up}(a). The Raman solver (RS) has pump powers and wavelengths as inputs and the gain profile is obtained at the output. We selected as an input optical data signal, a comb of ideal Nyquist-WDM channels packed at Nyquist limit, i.e. channel spacing equal to symbol-rate. This results in a flat optical spectrum (125 channels at 32 Gbaud covering the whole C-band). Channel power was set to 0 dBm, for a total WDM power of 21 dBm. Fiber parameters are listed in the following: span length $L_{span}=100$ km, attenuation $\alpha_S = 0.2$ dB/km for optical data signals and $\alpha_P =0.25$ dB/km for pumps, effective area: $A_{eff}=80$ $\mu$m$^2$, non-linear coefficient $\gamma$=1.26 1/W/km, chromatic dispersion $D=16.7$ ps/nm/km, and Raman coefficient $c_{R}=$0.4125 1/W/km.
Standard silica Raman efficiency profile has been assumed \cite{JLT_Curri_2016}.

For the generation of the data set, the following is performed: for $i=1,..,M$ draw $\lambda^i_1 \sim \mathcal{U}[1428,1463]$ nm, $\lambda^i_2 \sim \mathcal{U}[1463,1498]$ nm and ${P^i_1,P^i_2} \sim \mathcal{U}[0, 800]$ mW and run the Raman solver to obtain $[G_1^i,...,G_K^i]$. $\mathcal{U}$ denotes uniform distribution. Once the data set has been obtained, the data set $\mathcal{D}$ is formed and the multi-layer network, shown in Fig.~\ref{fig:system_set_up}(b), is employed to learn the mapping between $[G^i(\lambda_1),...,G^i(\lambda_{40}]$ and $[P^i_1,P^i_2,\lambda^i_1,\lambda^i_2)]$. 

Finally, to evaluate if the predicted pump powers and wavelengths result in the targeted gain profiles, the Raman solver is run using the predicted  pump powers and wavelengths. The corresponding predicted and targeted gain profiles are compared as shown in Fig.~\ref{fig:system_set_up}(c).

\section{Results and Validation}
\begin{figure}[t]
  \centering
  \includegraphics[width=8.1cm]{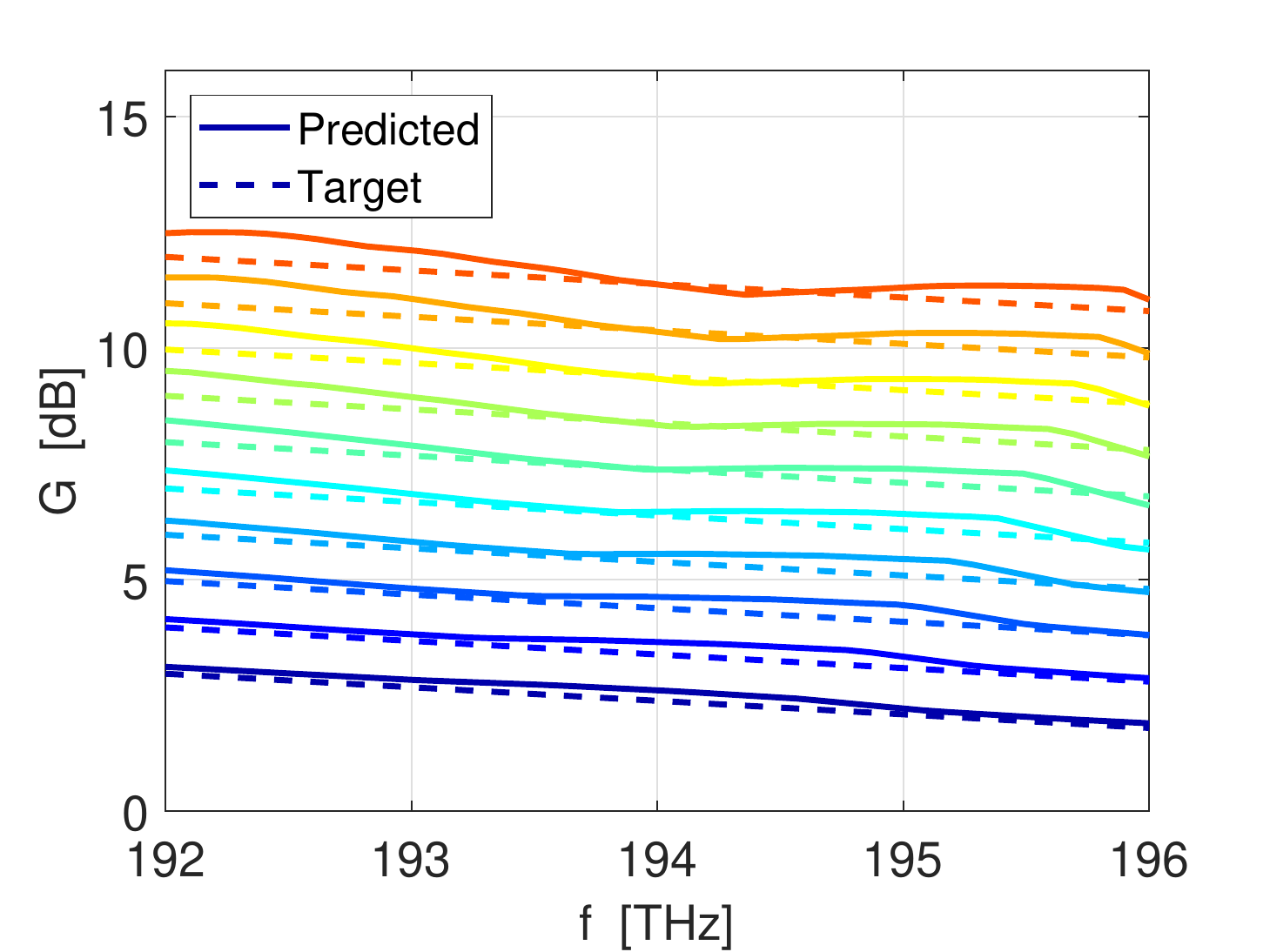}
  \includegraphics[width=8.1cm]{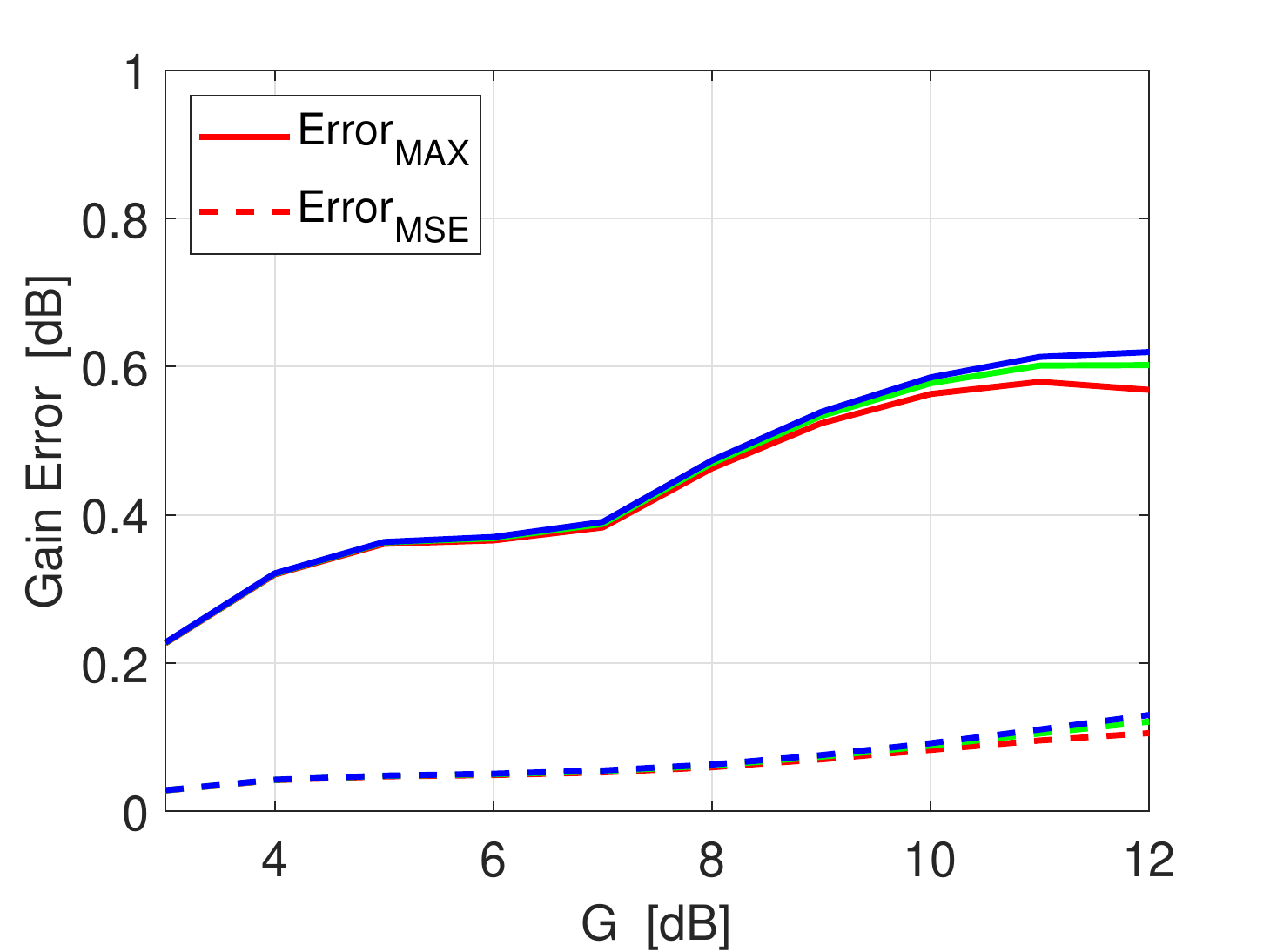}
\caption{\small{Validation for the tilted gain case. (a) Left: comparison between the predicted and target gain profiles. (b) Right: MSE and MAX gain error over the whole C-band as a function of gain level for different channel input powers (red: +3 dBm, green: 0 dBm and blue: -3 dBm).}}
\label{fig:validation_tilt}
\end{figure}

We start analyzing the ability of the multi-layer neural network to predict pump powers and wavelengths that would result in flat gain profiles. The considered gain levels range from 3 to 12 dB, i.e. the 60 \% of total span loss. This is a moderate pumping regime \cite{JLT_Curri_2016}, that covers all practical applications of RA in hybrid amplification and where RAs are most efficient.
In Fig.~\ref{fig:system_set_up}(c), predicted and target gain profiles are compared
when the input is a full C-band ideal Nyquist-WDM comb with 0 dBm per channel.
Good agreement is obtained and the maximum error is below 0.6 dB.\vspace{0.1cm} 

A second set of validations considers the same input signal but a tilted gain case: tilt is set to 1 dB over the considered C-band.
In Fig.~\ref{fig:validation_tilt}(a), we compare predicted and target gain profiles. The resulting predicted gain is in excellent agreement with the target gain profiles as the the maximum error is below 0.6 dB. In Fig.~\ref{fig:validation_tilt}(b), maximum gain error (Error$_{MAX}$) and mean square gain error (Error$_{MSE}$), evaluated over the whole bandwidth, are plotted as a function of gain for different input channel power levels (+3, 0 and -3 dBm). The Error$_{MAX}$ is limited to 0.6 dB and the Error$_{MSE}$ is much lower,
always below 0.2 dB.
Is is also observed that Error$_{MAX}$ and Error$_{MSE}$ increase as the gain levels become higher. 
In any case, maximum error is always limited to 0.6 dB and the MSE is much lower, always below 0.2 dB.
We also performed other tests varying the input WDM comb, reducing the number of channels
and placing them with different wavelength assignments, always finding a maximum error less than 0.6 dB.

\section{Conclusion}
A novel approach for the optimization of backward pumping Raman amplifiers, based on machine learning, has been presented. It has been demonstrated that a multi-layer neural network is able to accurately predict with low-complexity pump powers and wavelengths needed to obtain a targeted gain profile.
The proposed solution results in only 0.6 dB maximum error in an extensive set of validations over the whole C-band using only two pumps.

\small{
\subsection*{Acknowledgements}
This work is supported by the European Research Council
through the ERC-CoG FRECOM project (grant agreement
no. 771878).}

\end{document}